\documentclass[showpacs]{revtex4}
\usepackage{epsfig}
\def\beq{\begin{equation}}
\def\enq{\end{equation}}
\def\beqa{\begin{eqnarray}}
\def\enqa{\end{eqnarray}}
\def\MeV{\nobreak\,\mbox{MeV}}
\def\GeV{\nobreak\,\mbox{GeV}}
\def\keV{\nobreak\,\mbox{keV}}

\def\Tr{\mbox{ Tr}}

\def\G3{\lag g^3G^3\rag}
\def\pli{p^\prime}

\def\la{\lambda}

\def\ga{\gamma}
\def\Ga{\Gamma}
\def\om{\omega}

\def\si{\sigma}

\def\al{\alpha}

\def\lb{\label}

\newcommand{\rag}{\rangle}
\newcommand{\lag}{\langle}

\begin{document}

\title{\sc $X(3872)\rightarrow J/\psi\pi^+\pi^-$ and $X(3872)\to J/\psi
\pi^+\pi^-\pi^0$ decay widths from QCD sum rules}
\author{F.S. Navarra and M. Nielsen}
\affiliation{Instituto de F\'{\i}sica, Universidade de S\~{a}o Paulo, 
C.P. 66318, 05389-970 S\~{a}o Paulo, SP, Brazil}

\begin{abstract}

New spectroscopy from the B factories, the advent of CLEO-c and the BES 
upgrade renewed the interest in charmonia. Among the new measurements, the 
state  $X(3872)$ has received special attention due to its unexpected 
properties. Its structure has been studied with different theoretical  
approaches, most of them being able to  
reproduce the measured mass. A further test for the theoretical descriptions 
of the $X(3872)$ is to explain its narrow decay width. In this work we 
address the decays $X\to J/\psi\,\pi^+\pi^-\pi^0$ and  $X\to J/\psi\,\pi^+
\pi^-$, using QCD sum rules with the hypothesis that $X$ is a four quark 
state. 
\end{abstract}

\maketitle


During the last three years several new hadronic states have been observed, 
as for example the  $D_{sJ}^+(2317)$ \cite{dsexp} and the $X(3872)$ 
\cite{Xexpts}. 
The experimental observations were always followed by theoretical efforts to 
understand the basic properties of the new particles, in particular the mass 
and the decay width. In the charm sector, simple  potential models, which had 
been so successful in the past, failed in reproducing the masses of the new 
states. This was taken as an indication that these particles are not simple 
$q - \overline{q}$   bound states. As for the very narrow decay width, 
whereas it
is expected in the case of the $D_{sJ}^+(2317)$, since it decays through an  
isospin 
violating channel, in the case of the  $X(3872)$ it is really surprising. 
Even more
surprising is the observation, reported by the BELLE collaboration 
\cite{belleE}, 
that the $X$ decays to $J/\psi\,\pi^+\pi^-\pi^0$, with a strength that
is compatible to that of the  $J/\psi\pi^+\pi^-$ mode:
\beq
{Br(X\to J/\psi\,\pi^+\pi^-\pi^0)\over Br(X\to J/\psi\,\pi^+\pi^-)}
=1.0\pm0.4\pm0.3\;.
\label{data}
\enq 
This decay suggests 
an appreciable transition rate to $J/\psi\,\omega$ and establishes strong 
isospin violating effects. The measured  $X(3872)$  mass can be reproduced 
in several approaches and it is not yet
possible to discriminate between the different structures proposed for this 
state: tetraquark
\cite{Maiani,hrs},  cusp \cite{bugg}, hybrid \cite{li},  glueball
\cite{seth} or $D\bar D^*$ bound state \cite{tor,clopa,wong,pasu,ess}.

The theoretical study of the decay width  can help in clarifying this 
situation.   
In this work we use the  method of QCD  sum rules (QCDSR) \cite{svz,rry,SNB}
to study the hadronic decays of $X(3872)$ given in Eq.(\ref{data}),
considering $X$ as a four-quark state. 
In recent calculations \cite{sca,pec,x3872}, the
QCDSR approach was used to study the light scalar mesons, the 
$D_{sJ}^+(2317)$ meson and the $X(3872)$ meson
considered as four-quark states and a good agreement
with the experimental masses was obtained. In particular, in 
ref.\cite{x3872} we have considered the $X(3872)$ as the $J^{PC}=1^{++}$ 
state with the symmetric spin distribution: $[cq]_{S=1}[\bar{c}
\bar{q}]_{S=0}+[cq]_{S=0}[\bar{c}\bar{q}]_{S=1}$. The 
interpolating field for $X_q$ is given by:
\beq
j_\mu^q={i\epsilon_{abc}\epsilon_{dec}\over\sqrt{2}}[(q_a^TC\gamma_5c_b)
(\bar{q}_d\gamma_\mu C\bar{c}_e^T)+(q_a^TC\gamma_\mu c_b)
(\bar{q}_d\gamma_5C\bar{c}_e^T)]\;,
\label{field}
\enq
where $a,~b,~c,~...$ are colour indices, $C$ is the charge conjugation
matrix and 
$q$ represents the quark $u$ or $d$. 

As pointed out in \cite{Maiani}, isospin forbidden decays are possible if
$X$ is not a pure isospin state. Pure isospin states are:
\beq
X(I=0)={X_u+X_d\over\sqrt{2}},\;\;\;\mbox{and}\;\;\;\;X(I=1)={X_u-X_d\over
\sqrt{2}}.
\enq
If the physical states are just $X_u$ or $X_d$, the mass eigenstates,
maximal isospin violations are possible. Deviations from these two ideal
situations are described by a mixing angle between $X_u$ and $X_d$
\cite{Maiani}:
\beqa
X_l&=&X_u\cos{\theta}+X_d\sin{\theta},
\nonumber\\
X_h&=&-X_u\sin{\theta}+X_d\cos{\theta}.
\label{lh}
\enqa

In ref.~\cite{Maiani}, by considering the ratio of branching ratios given
in Eq.(\ref{data}), they arrived at $\theta\sim20^0$ and at
$\Gamma(X\to J/\psi\pi\pi)\sim5~\MeV$. However, to arrive at such small 
decay width they had to make a bold guess about the order of magnitude
of the $XJ/\psi V$ (where $V$ stands for the $\rho$ or $\omega$ vector 
meson) coupling constant: $g_{X\psi V}=0.475$. In this work we evaluate
the  $XJ/\psi V$ coupling constant directly from the QCD sum rules. 
For the light scalar mesons, considered as diquark-antidiquark states, the 
study of their vertex functions using the QCDSR approach was done 
in ref.\cite{sca}. The hadronic couplings determined in ref.\cite{sca} 
are consistent with existing experimental
data. In the case of the meson $D_{sJ}^+(2317)$ considered as a four-quark 
state, the QCDSR evaluation of the hadronic coupling constant $g_{D_{sJ}
D_s\pi}$ \cite{dsdpi} gives a partial decay width in the range $0.2\keV\leq
\Gamma(D_{sJ}^+(2317)\rightarrow D_s^+\pi^0)\leq40\keV$.

The QCDSR calculation for the vertex, $X(3872)J/\psi V$, centers
around the three-point function given by
\beq
\Pi_{\mu\nu\al}(p,\pli,q)=\int d^4x d^4y ~e^{i\pli.x}~e^{iq.y}
\Pi_{\mu\nu\al}(x,y),\mbox{ with }
\Pi_{\mu\nu\al}(x,y)=\lag 0 |T[j_\mu^{\psi}(x)j_{\nu}^{V}(y)
{j_\al^X}^\dagger(0)]|0\rag,
\lb{3po}
\enq
where $p=\pli+q$ and the interpolating fields 
are given by:
\beq
j_{\mu}^{\psi}=\bar{c}_a\gamma_\mu c_a,
\lb{psi}
\enq
\beq
j_{\nu}^{V}={N\over2}(\bar{u}_a\gamma_\nu u_a+(-1)^I\bar{d}_a\gamma_\nu 
d_a),
\lb{vec}
\enq
with $N=I=1$ for $V=\rho$, $N=1/3$, $I=0$ for $V=\om$ and
\beq
j_{\al}^{X}=aj_\al^u+bj_\al^d,
\lb{x}
\enq
where $j_\al^q$ is given in Eq.(\ref{field}) and
(see Eq.(\ref{lh})).
\beq
\mbox{for }X_l\left\{\begin{array}{l}
            a=\cos{\theta}\\
            b=\sin{\theta}\\
           \end{array}\right.,\;\;\;\;\;
\mbox{for }X_h\left\{\begin{array}{l}
            a=-\sin{\theta}\\
            b=\cos{\theta}\\
           \end{array}\right.
\label{ab}
\enq

Using the above definitions in Eq.(\ref{3po}) we arrive at
\beq
\Pi_{\mu\nu\al}(x,y)={-iN\over2\sqrt{2}}\left(a~\Pi^u_{\mu\nu\al}(x,
y)+(-1)^Ib~\Pi^d_{\mu\nu\al}(x,y)\right),
\label{piAI}
\enq
with
\beq
\Pi^q_{\mu\nu\al}(x,y)=\epsilon_{abc}\epsilon_{dec}\Tr\left[S^c_{ea'b}(x)
\gamma_\mu S^c_{a'b}(x)(\gamma_5C{S^{qT}_{b'a}}(y)C\gamma_\nu CS^{qT}_{db'}
(-y)C\gamma_\al-\gamma_\al CS^{qT}_{b'a}(y)C\gamma_\nu CS^{qT}_{db'}
(-y)C\gamma_5)\right],
\enq
where $S^q_{ab}(x-y)=\lag 0 |T[q_a(x)\bar{q}_b(y)]|0\rag$ is the full
quark $q$ propagator.

To evaluate the phenomenological side of the sum rule we  
insert, in Eq.(\ref{3po}), intermediate states for $X$, $J/\psi$ and $V$. 
Using the definitions: 
\beq
\lag 0 | j_\mu^\psi|J/\psi(\pli)\rag =m_\psi f_{\psi}\epsilon_\mu(\pli),\;\;
\;\;\lag 0 | j_\nu^V|V(q)\rag =m_{V}f_{V}\epsilon_\nu(q),\;\;\;\;
\lag X_q(p) | j_\al^q|0\rag =\la_q \epsilon_\al^*(p),
\lb{fp}
\enq
we obtain the following relation:
\beq
\Pi_{\mu\nu\al}^{q(phen)} (p,\pli,q)={\la_q m_{\psi}f_{\psi}m_Vf_{V}~
g_{X\psi V}(q^2)
\over(p^2-m_{X}^2)({\pli}^2-m_{\psi}^2)(q^2-m_V^2)}~\left(-\epsilon^{\al
\mu\nu\si}(\pli_\si+q_\si)-\epsilon^{\al\mu\si\ga}{\pli_\si q_\ga q_\nu
\over m_V^2}-\epsilon^{\al\nu\si\ga}{\pli_\si q_\ga\pli_\mu\over m_\psi^2}
\right) +\cdots\;,
\lb{phen}
\enq
where the dots stand for the contribution of all possible excited states, 
and the form factor, $g_{X\psi V}(q^2)$, is defined by the generalization 
of the on-mass-shell matrix element, $\lag J/\psi V|X\rag$,
for an off-shell $V$ meson: 
\beq
\lag J/\psi(\pli) V(q)|X(p)\rag=g_{X\psi V}(q^2)\epsilon^{\si\al\mu\nu}p_\si
\epsilon_\al(p)\epsilon_\mu^*(\pli)\epsilon_\nu^*(q),
\label{coup}
\enq
which can be extracted from the effective Lagrangian that describes the
coupling between two vector mesons and one axial vector meson \cite{Maiani}:
\beq
{\cal{L}}=ig_{X\psi V}\epsilon^{\mu\nu\al\si}(\partial_\mu X_\nu)\Psi_\al
V_\si.
\enq

From Eq.(\ref{phen}) we see that we have four independent structures in
the phenomenological side. For each one of these structures, $i$, we can 
write 
\beq
\Pi_i^{q(phen)}{A_i~
g_{X\psi V}(q^2)\over(q^2-m_V^2)(p^2-m_X^2)({\pli}^2-m_\psi^2)}
+\int_{4m_c^2}^\infty{\rho_i^{cont}(p^2,q^2,u)\over u-{\pli}^2}~du.
\label{mco}
\enq
In Eq.(\ref{mco}), $\rho_i^{cont}(p^2,q^2,u)$, gives the
continuum contributions, which can be 
parametrized as $\rho_i^{cont}(p^2,q^2,u)={b_i(u,q^2)\over s_0-p^2}
\Theta(u-u_0)$ \cite{dsdpi,io2,ennr}, with $s_0$ and $u_0$ 
being the continuum thresholds for $X$ and $J/\psi$ respectively.
Taking the limit $p^2={\pli}^2=-P^2$ and performing a
single Borel transformation to $P^2\rightarrow M^2$, we get ($Q^2=-q^2$): 
\beq
\Pi_i^{q(phen)}(M^2)= {A_i~g_{X\psi V}(Q^2)
\over(m_V^2+Q^2)(m_{X}^2-m_{\psi}^2)}\left(
e^{-m_\psi^2/M^2} -e^{-m_{X}^2/M^2}\right)+{B_i\over(m_V^2+Q^2)}~
e^{-s_0/M^2}+\int_{u_0}^\infty\rho_i^{cc}(u,Q^2)~e^{-u/M^2}du,
\label{paco}
\enq
where $B_i$ and $\rho_i^{cc}(u,Q^2)$ 
stand for the pole-continuum transitions and pure continuum contributions.
For simplicity, one assumes that the pure continuum contribution to the 
spectral density, $\rho_i^{cc}(u,Q^2)$, is given by the result obtained in 
the OPE side for the structure $i$. Asymptotic freedom ensures this 
equivalence for sufficiently large $u$.
Therefore, one uses the Ansatz: $\rho_i^{cc}(u,Q^2)=\rho_i^{OPE}(u,Q^2)$.
In Eq.(\ref{paco}), $B_i$ is a parameter which, together with the form 
factor, $g_{X\psi V}(Q^2)$, has to be determined from the sum rule.

\begin{figure}[h] \label{fig1}
\centerline{\epsfig{figure=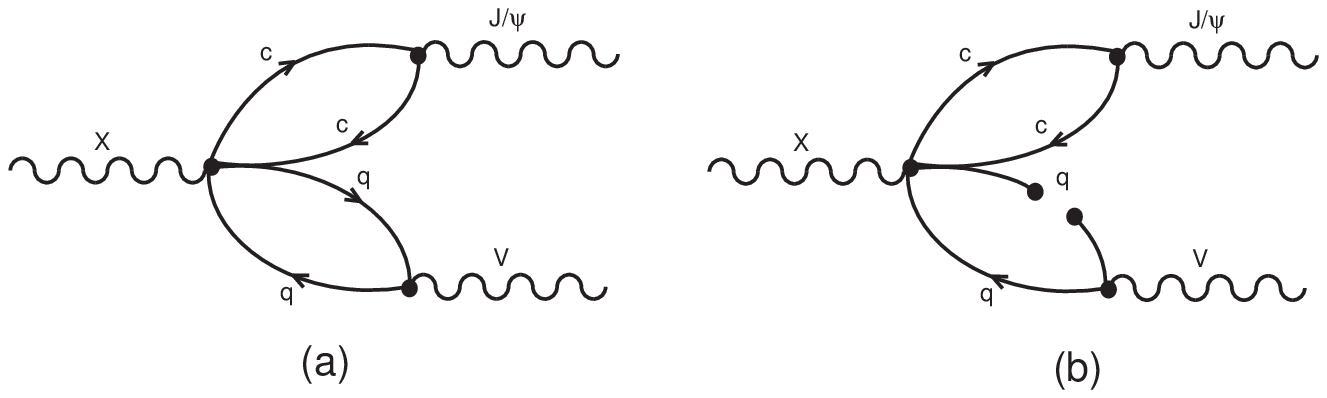,height=30mm}}
\centerline{\epsfig{figure=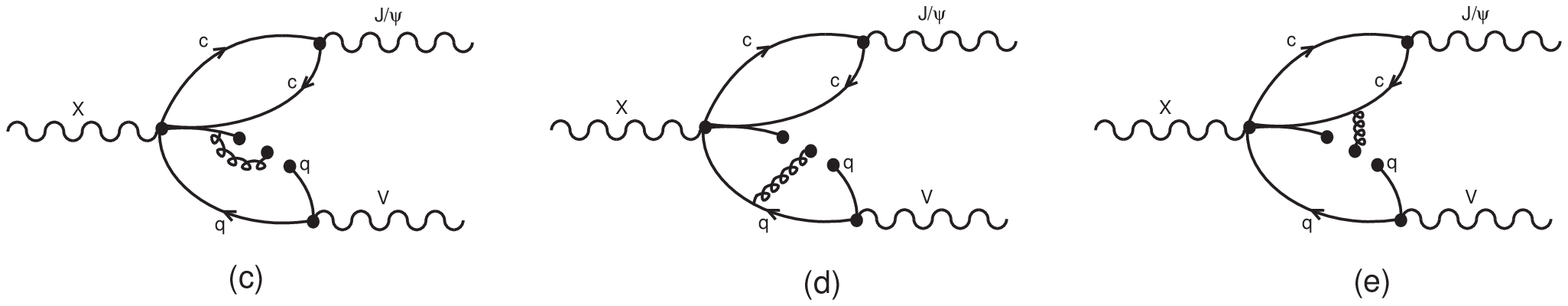,height=30mm}}
\caption{Diagrams which contribute to the OPE side of the sum rule.}
\end{figure} 
In the OPE side we work at leading order and consider the condensates up to 
dimension five, as shown in Fig.~1. To keep the charm quark mass finite, we
use the momentum-space expression for the charm quark propagator. We 
calculate the light quark part of the correlation
function in the coordinate-space, which is then Fourier transformed to the
momentum space in $D$ dimensions. The resulting light-quark part is 
combined 
with the charm-quark part before it is dimensionally regularized at $D=4$.
For each structure $i$,
we can write the Borel transform of the correlation function in 
the OPE side in terms of a dispersion relation:
\beq
\Pi_i^{q(OPE)}(M^2,Q^2)=\int_{4m_c^2}^\infty  \rho_i^{q(OPE)}(u,Q^2)
~e^{- u/M^2}du\;,
\lb{ope}
\enq
where the spectral density, $\rho_i^{q(OPE)}$, is given by the imaginary 
part of the correlation function. The perturbative term (diagram in Fig.~1(a)) 
contributes only to the structures $\epsilon^{\al\mu\nu\si}\pli_\si$ and 
$\epsilon^{\al\mu\si\ga}\pli_\si q_\ga q_\nu$, while the quark condensate
and mixed condensate (diagrams (b) to (e) in Fig.~1)
contribute to the structures $\epsilon^{\al\mu\nu\si}
q_\si$ and $\epsilon^{\al\nu\si\ga}\pli_\si q_\ga\pli_\mu$. Therefore,
to get more terms contributing in the OPE side we have two options for the
structures: $\epsilon^{\al\mu\nu\si}q_\si$ and $\epsilon^{\al\nu\si\ga}
\pli_\si q_\ga\pli_\mu$. In order to test the dependence of the results
with the chosen structure, we will work with these two structures.

Transferring the pure continuum contribution to the OPE side we get for
the structure  $\epsilon^{\al\nu\si\ga}\pli_\si q_\ga\pli_\mu$ (which we
call structure 1):
\beqa
\Pi_1^{q(OPE)}(M^2,Q^2)&=&{i\lag\bar{q}q\rag\over3\pi^2Q^2}\left[\left({m_0^2
\over3Q^2}-1\right)\int_{4m_c^2}^{u_0}du~e^{-u/M^2}~\sqrt{1-4m_c^2/u}
\left({1\over2}+{m_c^2
\over u}\right)+\right.
\nonumber\\
&-&\left.{m_0^2\over2^5}\int_0^1 d\al{1+3\al\over\al}~e^{-m_c^2\over
\al(1-\al)M^2}\right],
\label{est1}
\enqa
and for the structure  $\epsilon^{\al\mu\nu\si}q_\si$ (which we
call structure 2) we get:
\beqa
\Pi_2^{q(OPE)}(M^2,Q^2)&=&{i\lag\bar{q}q\rag\over3\pi^2Q^2}\left[\left({m_0^2
\over3Q^2}-1\right)\int_{4m_c^2}^{u_0}du~e^{-u/M^2}~u\sqrt{1-4m_c^2/u}
\left({1\over2}+{m_c^2\over u}\right)+\right.
\nonumber\\
&-&\left.{m_0^2m_c^2\over2^5}\int_0^1 d\al{1+3\al\over\al^2(1-\al)}
~e^{-m_c^2\over\al(1-\al)
M^2}\right].
\label{est2}
\enqa
In Eqs. (\ref{est1}) and (\ref{est2}) we have used the relation 
$\lag\bar{q}g\si.Gq\rag=m_0^2\lag\bar{q}q\rag$.

Making use of Eqs.~(\ref{ab}) and (\ref{piAI}), and working at the SU(2)
limit, {\it{i.e.}}, considering the quarks $u$ and $d$ degenerate, we 
arrive at three
sum rules for each structure, that can be written in the general expression:
\beq
C_i^{XV}(Q^2)\left(e^{-m_\psi^2/M^2}-e^{-m_X^2/M^2}\right)+B_i~e^{-s_0/M^2}=
-i{Q^2+m_V^2\over2\sqrt{2}}\Pi_i^{q(OPE)}(M^2,Q^2),
\label{sr}
\enq
where 
\beq
C_1^{XV}(Q^2)={f_\psi\over m_\psi}{\la_q\over m_X^2-m_\psi^2}A_{XV}(Q^2),
\mbox{ and }
C_2^{XV}(Q^2)={f_\psi m_\psi}{\la_q\over m_X^2-m_\psi^2}A_{XV}(Q^2),
\label{CXV}
\enq
and
\beqa
A_{X_l\rho}(Q^2)&=&m_\rho f_\rho{\cos{\theta}+\sin{\theta}\over \cos{\theta}-
\sin{\theta}}g_{X_l\psi\rho}(Q^2),
\nonumber\\
A_{X_h\rho}(Q^2)&=&-m_\rho f_\rho{\cos{\theta}-\sin{\theta}\over\cos{\theta}+
\sin{\theta}}g_{X_h\psi\rho}(Q^2),
\nonumber\\
A_{X_l\om}(Q^2)&=&A_{X_h\om}(Q^2)=3m_\om f_\om g_{X_{l,h}\psi\om}(Q^2).
\label{As}
\enqa

Since from Eq.(\ref{sr}) we see that the OPE side of the sum rule 
determines only one value for $C^{XV}$ for each structure (for a fixed
value of $Q^2$), we arrive at the following relations between the
form factors:
\beqa
{g_{X_{l}\psi\om}(Q^2)\over g_{X_{l}\psi\rho}(Q^2)}&=& {m_\rho f_\rho\over
3m_\om f_\om}{\cos{\theta}+\sin{\theta}\over \cos{\theta}-\sin{\theta}},
\nonumber\\
{g_{X_{h}\psi\om}(Q^2)\over g_{X_{h}\psi\rho}(Q^2)}&=&-{m_\rho f_\rho\over
3m_\om f_\om}{\cos{\theta}-\sin{\theta}\over \cos{\theta}+\sin{\theta}}.
\label{recoup}
\enqa

\begin{figure}[h] \label{fig1}
\centerline{\epsfig{figure=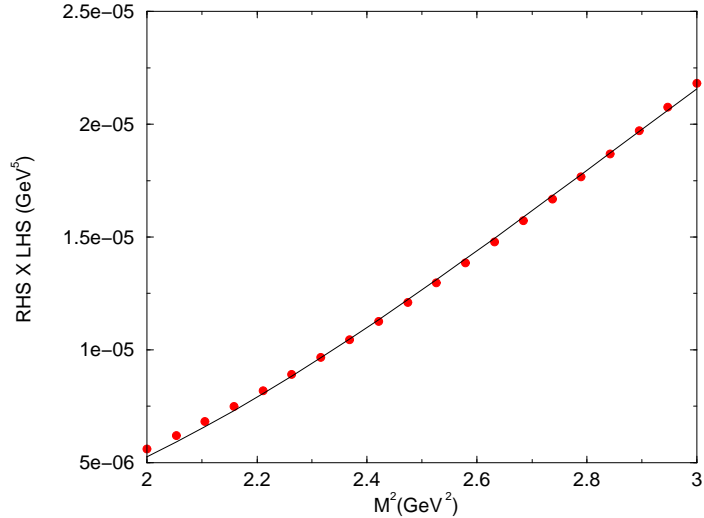,height=70mm}}
\caption{Dots: the RHS of Eq.(\ref{sr}), as a function of the Borel mass
for $Q^2=3~\GeV^2$.  
The solid line gives the fit of the QCDSR results through 
the LHS of Eq.(\ref{sr}) for $\Delta s_0=0.5~\GeV$.}
\end{figure} 

In the numerical analysis of the sum rules, the values used for the quark
masses  and condensates are: 
$m_c=1.2\,\GeV$, $\lag\bar{q}q\rag=\,-(0.23)^3\,\GeV^3$,
$m_0^2=0.8\,\GeV^2$. For the meson parameters we use their experimental 
values \cite{pdg}: $m_\rho=0.776\GeV$, $m_\om=0.782\GeV$, $m_\psi=3.1\GeV$, 
$m_X=3.872\GeV$, $f_\rho=0.157\GeV$, $f_\om=0.046\GeV$ and $f_\psi=0.405
\GeV$.
We evaluate our sum rules in the range $2.0 \leq M^2 \leq 3.0$, 
which is the range where the two-point function for $X(3872)$ shows good
OPE convergence and where the pole contribution is bigger than the
continuum contribution \cite{x3872}. We also use three different
values for $s_0 = (3.872 + \Delta s_0)^2\GeV^2$: $\Delta s_0 = 0.4$ GeV,
$\Delta s_0 = 0.5$ GeV and $\Delta s_0 = 0.6$ GeV. For $u_0$ we use
$u_0=(m_\psi+0.5)^2\GeV^2$. The meson-current coupling, $\la_q$, defined in 
Eq.(\ref{fp}), can be determined from the two-point sum rule \cite{x3872}.
In Table I we give the results obtained from ref.\cite{x3872} for three
different values of $s_0$.
\begin{center}
\small{{\bf Table I:} Numerical results for the meson-current coupling}
\\
\vskip3mm

\begin{tabular}{|c|c|}  \hline
$\la_q~(\GeV^5)$ & $\Delta s_0~(\GeV)$ \\
\hline
 $(1.85\pm0.01)\times10^{-2}$ &0.4 \\
\hline
$(1.94\pm0.03)\times10^{-2}$ &0.5 \\
\hline
$(2.02\pm0.06)\times10^{-2}$ & 0.6 \\
\hline
\end{tabular}\end{center}

We start with the structure 1.
In Fig.~2 we show, through the circles, the right-hand side (RHS) of 
Eq.(\ref{sr}) for $Q^2=3\GeV^2$, as a function of the Borel mass.

To determine $g_{X\psi V}(Q^2)$ we fit the QCDSR results with the analytical
expression in the left-hand side (LHS) of Eq.(\ref{sr}),
and we get (using $\Delta s_0=0.5~\GeV$): $C_1^{XV}(Q^2=3\GeV^2)=7.01
\times10^{-4}~\GeV^5$ and $B_1=-1.28\times10^{-3}~\GeV^5$. Using the 
definition of $C_1^{XV}(Q^2)$ in Eq.(\ref{CXV})
we get $A_{XV}(Q^2=3\GeV^2)=1.49~\GeV^2$. Allowing $Q^2$ to vary in the
interval $2.5\leq Q^2\leq4.5~\GeV^2$,  we show, in Fig.~3, through the 
circles, the momentum dependence of $A_{XV}(Q^2)$.

From Eq.(\ref{As}), we see that all  form factors are  related with
the function $A_{XV}(Q^2)$. Since the coupling constant is
defined as the value of the form factor at the meson pole: $Q^2=-m_V^2$,
to determine the coupling constant we have to extrapolate the QCDSR results
to a $Q^2$ region where the sum rules are no longer valid (since the QCDSR 
results are valid in the deep Euclidian region). To do that we parametrize
the QCDSR results through a analytical form.
In Fig.~3 we also show that the $Q^2$ dependence of $A_{XV}(Q^2)$ can be
well reproduced by the monopole parametrization (solid line):
\beq
A_{XV}(Q^2)= {66.8\over Q^2+41.8}\;,
\label{mo}
\enq
from where we can extract the value of $A_{XV}(Q^2)$ at the meson pole:
$A_{XV}(Q^2=-m_V^2)=1.62~\GeV^2$.

\begin{figure}[h]  \label{fig2}
\centerline{\epsfig{figure=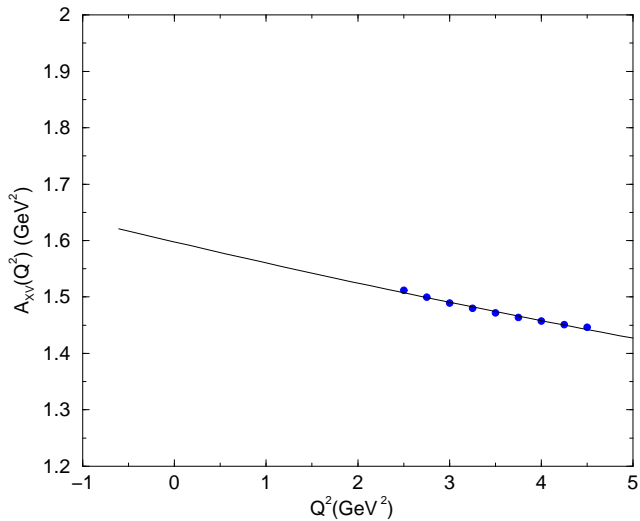,height=70mm}}
\caption{Momentum dependence of  $A_{XV}$ for
$\Delta s_0=0.5\,\GeV$. The solid line gives the 
parametrization of the QCDSR results (circles) through Eq.~(\ref{mo}).}
\end{figure}

Doing the same kind of analysis for the other values of the continuum  
threshold we show, in Table II, the monople parametrizations of the QCDSR
results, as well as their values at the off-shell meson pole.

\begin{center}
\small{{\bf Table II:} Monopole parametrization of the QCDSR results for
the structure 1, for different values of $\Delta s_0$}
\\
\vskip3mm

\begin{tabular}{|c|c|c|}  \hline
$\Delta s_0~(\GeV)$&$A_{XV}(Q^2)~(\GeV^2)$ & $A_{XV}(Q^2=-m_V^2)~(\GeV^2)$\\
\hline
 0.4 & ${70.2\over Q^2+41.6}$ & 1.71 \\
\hline
 0.5 & ${66.8\over Q^2+41.8}$ & 1.62 \\
\hline
 0.6 & ${63.8\over Q^2+41.7}$ & 1.55 \\
\hline
\end{tabular}\end{center}

\begin{figure}[h] \label{fig3}
\centerline{\epsfig{figure=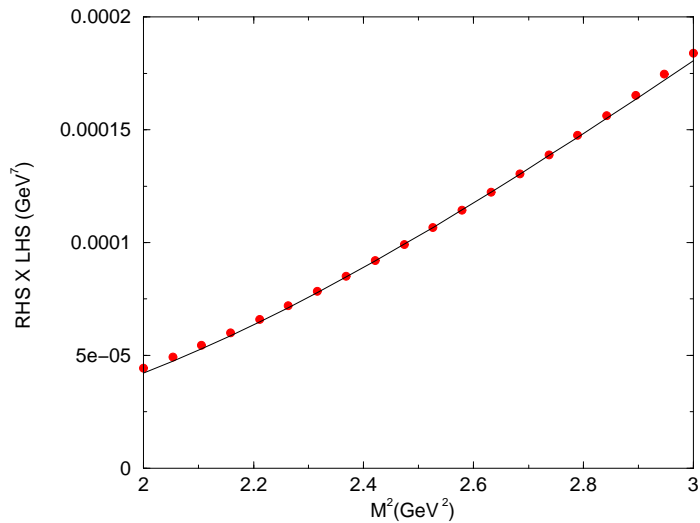,height=70mm}}
\caption{Dots: the RHS of Eq.(\ref{sr}), for the structure 2, as a function 
of the Borel mass for $Q^2=3~\GeV^2$.  
The solid line gives the fit of the QCDSR results through 
the LHS of Eq.(\ref{sr}) for $\Delta s_0=0.4~\GeV$.}
\end{figure} 

In the case of the structure 2, the RHS of  Eq.(\ref{sr}) can also be very 
well parametrized with the analytical expression in the LHS of Eq.(\ref{sr}),
as can be seen in Fig.~4.
We get (using $\Delta s_0=0.4~\GeV$): $C_2^{XV}(Q^2=3\GeV^2)=5.56
\times10^{-3}~\GeV^7$ and $B_2=-3.46\times10^{-3}~\GeV^7$. Using the 
definition of $C_2^{XV}(Q^2)$ in Eq.(\ref{CXV})
we get $A_{XV}(Q^2=3\GeV^2)=1.29~\GeV^2$. The $Q^2$ behaviour of
$A_{XV}(Q^2)$ can also be well represented by a monopole form in the case 
of structure 2, with  a precision similar to the one shown in Fig.~3.
In Table III we give the monople parametrizations of the QCDSR results for 
the structure 2, as well as their values at the off-shell meson pole.

\begin{center}
\small{{\bf Table III:} Monopole parametrization of the QCDSR results for
the structure 2, for different values of $\Delta s_0$}
\\
\vskip3mm

\begin{tabular}{|c|c|c|}  \hline
$\Delta s_0~(\GeV)$&$A_{XV}(Q^2)~(\GeV^2)$ & $A_{XV}(Q^2=-m_V^2)~(\GeV^2)$\\
\hline
 0.4 & ${59.0\over Q^2+42.6}$ & 1.45 \\
\hline
 0.5 & ${56.0\over Q^2+42.8}$ & 1.34 \\
\hline
 0.6 & ${54.2\over Q^2+42.9}$ & 1.28 \\
\hline
\end{tabular}\end{center}

Comparing the results in Tables II and III we see that, althought the
results from the structure 2 are somewhat smaller than the results from
the structure 1, they are still compatible with each other. We will use 
these differences to estimate the uncertainties in our results.

From Eq.~(\ref{As}) we see that, in the case of the meson $\om$, there is
no mixing angle dependence in the relation between $A_{XV}$ and 
$g_{X\psi \om}$. Therefore we can use the results in Tables II and III to 
directly estimate the $XJ/\psi\om$ coupling constant. We get 
\beq
g_{X\psi \om}=13.8\pm2.0,
\label{gome}
\enq
which is much bigger than the guess made in ref.\cite{Maiani}:
$g_{X\psi V}=0.475$.

Having the coupling constant and the relations in Eqs.~(\ref{As}) and 
(\ref{recoup}), we can estimate the decay widths of the processes
$X\to J/\psi\,\pi^+\pi^-\pi^0$ and $X\to J/\psi\,\pi^+\pi^-$ by supposing 
that the $2\pi$ and $3\pi$ decays are dominated by the $\rho$ and
$\om$ vector mesons respectively. In the narrow width approximation
we have:
\beq
{d\Gamma\over ds}(X\to J/\psi (n\pi))={1\over8\pi m_X^2}|{\cal{M}}|^2{m_X^2-
m_\psi^2+s\over2m_X^2}~{\Ga_V m_V\over\pi}{p(s)\over(s-m_V^2)^2+
(m_V\Ga_V)^2}B_{V\to n\pi},
\label{de1}
\enq
with $n=2,3$ for $V=\rho,~\om$. In Eq.(\ref{de1}),
$s$ is the invariant mass-squared of the pions, $\Ga_V$ and
$B_{V\to n\pi}$ are, respectively, the total decay  width and the branching 
ratio of the $V\to n\pi$ decay. The decay momentum $p(s)$ is given by
\beq
p(s)={\sqrt{\la(m_X^2,m_\psi^2,s)}\over2m_X},
\enq
with $\la(a,b,c)=a^2+b^2+c^2-2ab-2ac-2bc$. 

The invariant amplitude squared can be obtained from the matrix element in
Eq.(\ref{coup}). We get:
\beq
|{\cal M}|^2={g_{X\psi V}^2\over3}\left(4m_X^2-{m_\psi^2+s\over2}
+{(m_X^2-m_\psi^2)^2\over2s}+{(m_X^2-s)^2\over2m_\psi^2}\right),
\label{m2}
\enq
where we have replaced the form factor, $g_{X\psi V}(s)$ by the coupling
constant $g_{X\psi V}$, since from Tables II and III we can see that the
form factor is very flat over the region, $(n~m_\pi)^2\leq s\leq(m_X-
m_\psi)^2$, over which Eq.(\ref{de1}) will be integrated. Using the relations
between the coupling constants from Eq.(\ref{recoup}) we get the following
relations between the decay widths:
\beqa
\left({\Ga(X_l\to J/\psi~3\pi)\over\Ga(X_l\to J/\psi~2\pi)}\right)&=&
{m_\rho^2f_\rho^2\over9m_\om^2f\om^2}\left(\cos{\theta}+\sin{\theta}\over
\cos{\theta}-\sin{\theta}\right)^2{I_\om\over I_\rho},
\nonumber\\
\left({\Ga(X_h\to J/\psi~3\pi)\over\Ga(X_h\to J/\psi~2\pi)}\right)&=&
{m_\rho^2f_\rho^2\over9m_\om^2f\om^2}\left(\cos{\theta}-\sin{\theta}\over
\cos{\theta}+\sin{\theta}\right)^2{I_\om\over I_\rho},
\label{rela}
\enqa
where we have defined
\beqa
I_V&=&{\Ga_Vm_V\over\pi}\int_{(n~m_\pi)^2}^{(m_X-m_\psi)^2} ds\left[ 
\left(4m_X^2-{m_\psi^2+s\over2}
+{(m_X^2-m_\psi^2)^2\over2s}+{(m_X^2-s)^2\over2m_\psi^2}\right)
\times\right.
\nonumber\\
&\times&\left.{m_X^2-m_\psi^2+s\over2m_X^2}{p(s)\over(s-m_V^2)^2+
(m_V\Ga_V)^2}\right]B_{V\to n\pi}.
\enqa

Since the relations in Eq.(\ref{rela}) do not depend on the value of
the coupling constant we get
\beq
\left({\Ga(X_{l,h}\to J/\psi~3\pi)\over\Ga(X_{l,h}\to J/\psi~2\pi)}\right)
=0.152\left(\cos{\theta}\pm\sin{\theta}\over
\cos{\theta}\mp\sin{\theta}\right)^2.
\enq
Therefore, using the central experimental data given in Eq.(\ref{data}) we 
obtain for the mixing angle
\beq
\theta\simeq\pm23.5^0
\label{mix}
\enq
for $X_l$ or $X_h$ respectively, which is in agreement with the result 
obtained in \cite{Maiani}: $\theta\simeq\pm20^0$.

Since, with the determination of the mixing angle in Eq.(\ref{mix}) by
imposing the ratio in Eq.(\ref{data}), we obtain the same width for any
of the four decays in Eq.(\ref{rela}), we can use the value of the coupling 
constant determined in Eq.(\ref{gome}) to evaluate the partial decay width. 
We get
\beq
\Ga(X\to J/\psi ~(n\pi))=(50\pm15)~\MeV,
\label{width}
\enq
which is much bigger than the experimental total width: $\Ga(X(3872))
<2.3~\MeV$. 

As a matter of fact, a large partial decay width was expected in this case.
The initial state already contains all the four quarks needed for the decay, 
and there is no violating rules prohibiting the decay. Therefore, the decay
is allowed as in the case of the light scalars $\sigma$ and $\kappa$
studied in \cite{sca}, which widths are of the order of 400 MeV. 
However,
even when there is no violating rules prohibiting the decay,
the decay can be prevented due to a non-trivial color structure in the
initial state.
In ref.\cite{ennr}, an alternative technique was developed to obtain the
form factor and coupling constant for multiquark particles. By multiquark
we mean that the initial state contains the same number of valence quarks
as the number of valence quarks in the final states. In this case, as can be
seen in Fig.~1, the generic decay diagram in terms of quarks has two
``petals'', one associated with the $J/\psi$ and the other with the other
vector meson $V$. Among the diagrams in Fig.~1 there are two distinct
subsets. In the first (diagrams from (a) to (d)) there is no gluon line 
connecting
the petals and, therefore, no color exchange between the two final mesons
in the decay. A diagram of this type was called color-disconected (CD) 
diagrams in ref.~\cite{ennr}. If there
is no color exchange, the final state containing two color singlets was
already present in the initial state. In this case the tetraquark had 
a component similar to a $J/\psi-V$ molecule. The other subset of diagrams 
is represented by diagram in Fig.~1(e), where there is a color exchange 
between the 
petals.  This type of diagram represents the case where the $X$ is a 
genuine four-quark state with a complicated color structure. These diagrams 
are called color-conected (CC). In our approach we have considered all 
kinds of diagrams. However, if we consider only 
the CC diagrams, which means considering only the diagram (e) in Fig.~1,
we get $g_{X\psi \om}^{(CC)}=1.6\pm0.3$, 
and therefore
\beq
\Ga_{CC}(X\to J/\psi ~(n\pi))=(0.7\pm0.2)~\MeV.
\label{widthcc}
\enq

This procedure may appear somewhat unjustified. However, we do believe that
there should be a particular choice of the interpolating field, which 
represents a genuine four-quark state, for which CD diagrams vanish.
From our calculation we find out that the interpolating field in 
Eq.(\ref{field}) has a component similar to a $J/\psi-V$ molecule.

To summarize: we have presented a QCD sum rule study of the three--point 
functions of 
the hadronic decays of $X(3872)$ meson, considered
as a diquark antidiquark four quark state. 
Supposing that the physical state is a mixture between
the isospin eigenstates, we find that the QCD sum rules result
for the mixing angle is compatible with the result found in \cite{Maiani}. 
However, we get a partial decay width much bigger than the experimental 
total decay width. Therefore, we conclude that our particular choice
of the interpolating field has a $J/\psi V$ molecule component, and 
is not the most appropriate candidate to explain the very small width of 
the meson $X(3872)$. Further studies, using different interpolating fields, 
are necessary for a better understanding of the structure of the meson 
$X(3872)$.

\end{document}